\documentclass[12pt]{article}
\usepackage{latexsym}
\usepackage{amsmath}
\usepackage{amssymb}
\usepackage{epsfig}
\usepackage{epic}

\numberwithin{equation}{section}
\numberwithin{figure}{subsection}

\begin{document}

\def\k{\kappa}
\def\half{\fraction{1}{2}}
\def\fraction#1#2{ { \scriptstyle \frac{#1}{#2} }}
\def\der#1{\frac{\partial}{\partial #1}}
\def\d{\partial}
\def\p2{\fraction{\pi}{2}}
\def\s{\sigma}
\def\Id{|I\rangle}
\def\Tr{\mathrm{Tr}}
\def\z{\xi}
\def\eq#1{eq.(\ref{eq:#1})}
\def\Eq#1{Eq.(\ref{eq:#1})}
\def\L{\mathcal{L}}
\def\B{\mathcal{B}}
\def\Langle{\langle\langle}
\def\Rangle{\rangle\rangle}

\begin{titlepage}
\rightline{\today}

\begin{center}
\vskip 1.0cm
{\Large \bf{Tachyon Vacuum in Cubic Superstring 
Field Theory}}
\\
\vskip 1.0cm

{\large Theodore Erler}
\vskip 1.0cm
{\it {Harish-Chandra Research Institute} \\ 
{Chhatnag Road, Jhunsi, Allahabad 211019, India}}
\\ E-mail:terler@mri.ernet.in \\

\vskip 1.0cm
{\bf Abstract}
\end{center}

In this paper we give an exact analytic solution for tachyon condensation 
in the modified (picture 0) cubic superstring field theory. We prove the 
absence of cohomology and, crucially, reproduce the correct 
value for the D-brane tension. The solution is surprising for two reasons: 
First, the existence of a tachyon vacuum in this theory has not been 
definitively established in the level expansion. Second, the solution 
{\it vanishes} in the GSO$(-)$ sector, implying a ``tachyon vacuum'' 
solution exists even for a {\it BPS} D-brane.

\noindent

\medskip

\end{titlepage}

\newpage

\baselineskip=18pt

\tableofcontents

\section{Introduction}

Since Schnabl provided\cite{Schnabl} an analytic solution for tachyon 
condensation\cite{Sen} in bosonic open string field theory, it has been of some
interest to extend the analysis to superstrings. The most robust approach 
would utilize Berkovits's WZW-type superstring field theory\cite{Berkovits}, 
the only superstring framework for which reliable evidence for Sen's 
conjectures is available\cite{Berk_trunc}. However, the 
nonpolynomial structure of the WZW action makes it a challenge to identify a 
solution and compute the brane tension analytically. Thus it seems worth 
reconsidering an old, if somewhat questionable proposal, which formulates 
superstring field theory as a cubic action for a picture number 0 string 
field---the so-called modified cubic superstring field 
theory\cite{Russians} (see refs.\cite{Raeymaekers,Ohmori} for reviews). The 
action is,
\begin{equation}S = \frac{1}{2}\langle\langle\Psi,Q_B\Psi\rangle\rangle
+\frac{1}{3}\langle\langle\Psi,\Psi*\Psi\rangle\rangle
\end{equation} 
where $\Psi$ is a ghost number 1, picture number 0 string field in the small 
Hilbert space of the matter+ghost superconformal field theory 
$X,\psi,b,c,\xi,\eta,\phi$. The only qualitative difference from the bosonic 
string is the definition of the bracket $\Langle,\Rangle$, 
which requires insertions of two inverse picture changing operators at the 
open string midpoint. As a correlator in the upper half plane,
\begin{equation}\Langle \Psi,\Phi \Rangle = \langle Y_{-2}\,I\circ\psi(0)\,
\phi(0)\rangle_{UHP}\ \ \ \ \ I(z)=-\frac{1}{z}\end{equation}
where (using the doubling trick)\footnote{There are actually many possible 
choices for $Y_{-2}$, defining inequivalent string field theories off-shell. 
We will stick with the definition \eq{Y-2} since it the most canonical 
choice.},
\begin{equation}Y_{-2}=Y(i)Y(-i)\ \ \ \ \ \ 
Y(z)=-\d\xi e^{-2\phi}c(z)\label{eq:Y-2}\end{equation}
Since $Y(z)$ is dimension 0, BRST invariant and inserted at the midpoint, one 
can easily verify that all of the usual Chern-Simons like axioms are satisfied.

Though this action is very simple, as yet it is uncertain whether it defines 
an acceptable string field theory. One well-known objection\cite{Berkovits3} 
is that $Y_{-2}$ has a nontrivial kernel, so the expected cubic equations of 
motion are reproduced only up to terms annihilated by $Y_{-2}$. However, the 
offending fields would be very singular at the string midpoint\footnote{The
subalgebra of wedge states and related operators relevant for analytic 
calculations does not naturally produce fields in the kernel of $Y_{-2}$.
On the other hand, it is difficult to study vacuum string field 
theory\cite{Ohmori2} with the canonical kinetic operator $\sim c(i)$ in this 
framework.}, so it is unclear at what level this phenomenon will cause 
problems. Perhaps the 
ultimate test is to see whether the action reproduces the expected 
physics of tachyon condensation. Unfortunately, the answer is 
unclear\cite{Raeymaekers,Russians2,Ohmori2}. A candidate vacuum solution has 
been identified at the first few levels, but---as can be seen in figure 
\ref{fig:level}---the energy does not appear to converge. In fact, at level 
$(\frac{5}{2},5)$ an odd thing happens: the tachyon effective potential 
hits a singularity before the stable vacuum is reached, meaning that the 
(conjectured) nonperturbative vacuum lies on disconnected branch of the 
potential\cite{Raeymaekers}. Despite these oddities, the striking thing about 
the energies in fig.\ref{fig:level} is that they are so close to 
the right answer. It is hard to believe this is a coincidence, but certainly 
more computations would be necessary to establish some sort of convergence.

\begin{figure}
\begin{center}
\begin{tabular}{|c|c|c|c|c|c|}\hline 
Level: & (0,0) & ($\half$,1) & (2,4) & (2,6) & ($\frac{5}{2}$,5) \\
\hline
Percent Brane Tension & --- & 97\% & 108\% & 99\% & 91\% \\ \hline 
\end{tabular}
\end{center}
\caption{\label{fig:level} Percent of brane tension produced for the 
tachyon vacuum in the modified cubic theory at various levels. Results taken 
from Raeymaekers\cite{Raeymaekers} and Ohmori\cite{Ohmori2}.}
\end{figure}

In this paper we study the modified cubic theory from an analytic 
perspective, finding that---despite the above problems---the theory has 
a solution which can be interpreted as the endpoint of tachyon condensation.
The crucial component is the calculation of the correct brane tension, which 
serves as a successful test of the cubic action. Surprisingly, the solution 
{\it vanishes} in the GSO$(-)$ sector, implying that the vacuum exists even 
for the field theory on a BPS brane. 

This paper is organized as follows. In section \ref{sec:solution} we give the
algebraic setup and present the solution. We give a careful discussion of 
the ``$\psi_N$ piece'' which requires some additional modification to 
reproduce the correct brane tension. In section \ref{sec:Energy} we 
evaluate the energy. Due to the extraordinary simplicity of the crucial 
correlator, the calculation is very easy---much simpler than for the bosonic 
string. In fact, we are even able to compute the energy directly in the 
$\L_0$ level expansion. We end with some conclusions.

\section{Solution}
\label{sec:solution}

We seek a generalization of Schnabl's solution for the cubic superstring
equations of motion. The first step is identifying the relevant worldsheet 
degrees of freedom for expressing the solution. In split string 
notation\cite{Okawa,Erler1,Erler2}, we claim these degrees of freedom are given
by four string fields,
\begin{eqnarray}K &=& \mathrm{Grassmann\ even,\ gh}\#= 0\nonumber\\ 
B &=& \mathrm{Grassmann\ odd,\ gh}\#= -1\nonumber\\
c &=& \mathrm{Grassmann\ odd,\ gh}\#= 1\nonumber\\
\gamma^2 &=& \mathrm{Grassmann\ even,\ gh}\#= 1
\end{eqnarray}
defined,
\begin{eqnarray}
K &=& -\frac{\pi}{2}(K_1)_L\Id\ \ \ \ \ K_1 = L_1+L_{-1}\nonumber\\ 
B &=& -\frac{\pi}{2}(B_1)_L\Id\ \ \ \ \ B_1 = b_1+b_{-1}\nonumber\\ 
c &=& -\frac{1}{\pi}c(1)\Id\nonumber\\
\gamma^2 &=& \frac{1}{\pi}\gamma^2(1)\Id\ \ \ \ \ 
\gamma^2(z)=\eta\d \eta e^{2\phi}(z)
\label{eq:KBJ}\end{eqnarray}
where $\Id$ is the identity string field and the subscript $L$ denotes taking
the left half of the corresponding charge (integrating the current 
counter-clockwise on the positive half of the unit circle). These fields 
satisfy the algebraic relations,
\begin{eqnarray}[B,K]=0\ &\ &\ [B,\gamma^2]=0\ \ \ \ \ \ 
[c,\gamma^2]=0\nonumber\\
\ [B,c]=1\ &\ & B^2=c^2=0 \label{eq:alg}
\end{eqnarray}
and have BRST variations ($d=Q_B$),
\begin{eqnarray}dc &=& cKc + \gamma^2\ \ \ \ \ \ \ \ \ \ \ \ \ \ \ 
dB=K\nonumber\\
d\gamma^2 &=& cK\gamma^2-\gamma^2Kc\ \ \ \ \ \ \ \ \ dK=0
\label{eq:BRST}\end{eqnarray}
As another bit of notation, we denote
\begin{equation}F=e^{K/2}=\Omega^{1/2}\end{equation}
for the square root of the $SL(2,\mathbb{R})$ vacuum $\Omega=e^K$.

With these preparations, the conjectured vacuum solution to the cubic 
equations of motion
\begin{equation}d\Psi +\Psi^2=0\end{equation} 
is
\begin{equation}\Psi = Fc\frac{KB}{1-F^2}cF - FB\gamma^2F\label{eq:super_sol}
\end{equation}
We recognize the first term as Schnabl's solution for the bosonic 
string; the second term is a surprisingly simple superstring ``correction.''
The solution is real and satisfies the Schnabl gauge $\B_0\Psi=0$.
There are many ways of ``deriving'' \eq{super_sol}, but perhaps the simplest 
is to translate Okawa's pure gauge form\cite{Okawa} using the modified BRST 
identities \eq{BRST}. Following Ellwood and Schnabl\cite{cohomology}, 
the proof of absence of cohomology is immediate. We simply note the 
existence of a homotopy operator $A$ satisfying,
\begin{equation}d_{\Psi}A = dA + [\Psi,A] = 1\end{equation}
The homotopy operator is the same as for the bosonic string,
\begin{equation}A = -B\int_0^1dt \Omega^t \label{eq:A}\end{equation} 
since the $B$ kills the correction term $-FB\gamma^2F$ and the rest of the 
computation reduces to the bosonic derivation.

Perhaps the most alarming aspect of the solution \eq{super_sol} is the absence
of GSO$(-)$ states. In particular, the solution exists even for 
the field theory on a {\it BPS} D-brane. While this is quite counterintuitive,
we can offer some insight as to why this is possible, at least at the 
mathematical level. Note that, in some sense, the modified cubic theory has
{\it two} tachyons: the physical tachyon in the GSO$(-)$ sector, corresponding
to the vertex operator $\gamma(0)$; and the ``auxiliary tachyon'' in the 
GSO$(+)$ sector, corresponding to the vertex operator $c(0)$. The auxiliary 
tachyon does not represent a physical instability since $c(0)$ cannot be 
placed on shell. Nevertheless, the condensation of $c(0)$ is really what's 
responsible for the absence of open strings at the vacuum. One way of seeing 
this is through vacuum string field theory\cite{VSFT}, which can be obtained 
from \eq{super_sol} after an infinite reparameterization in the $\L_0$ level 
expansion. At the first two $\L_0$ levels the solution is,
\begin{equation}\Psi = -FcF +\left(\frac{1}{2}FcKBcF- FB\gamma^2F\right)+...
\end{equation}
Now perform an infinite reparameterization of the form discussed in 
refs.\cite{Erler1,RZO},
\begin{equation}\Psi\ \ \rightarrow\ \ \Psi_\alpha = 
\exp\left[\frac{1}{2}\ln\alpha(\L_0-\L_0^*) \right]\Psi\end{equation}
with $\alpha\to 0$. To leading order, the solution becomes
\begin{equation}\Psi_\alpha = -\frac{1}{\alpha}c +... 
\mathcal{O}(\alpha^0)\end{equation}
and the corresponding kinetic operator is,
\begin{equation}d_\Psi\ \ \rightarrow\ \ d_{\Psi_\alpha} 
= -\frac{1}{\pi\alpha}(c(1)-c(-1))+...\mathcal{O}(\alpha^0)\end{equation}
This is just the kinetic operator for (a form 
of\footnote{This reparameterization squeezes towards the endpoints rather than
the midpoint, which is why we obtain $c(1)$ rather than $c(i)$. The author
thanks E. Fuchs for a useful discussion on this limit.}) vacuum string field 
theory. If the solution had some expectation value for the tachyon 
$\gamma(0)$, this would have appeared as a subleading divergence 
$\alpha^{-1/2}$ in the vacuum kinetic operator. Such terms can be 
accommodated into the vacuum string field theory framework\cite{Ohmori1}, but 
really it is the leading divergence from the $c$ ghost which is responsible 
for the absence of cohomology. Note that these 
comments naively apply to the Berkovits theory as well\footnote{In fact, 
a GSO$(+)$ vacuum solution to the Berkovits theory has already been 
conjectured in ref.\cite{Fuchs_super}. The relation to our cubic solution is 
$e^{-\Phi}de^\Phi=\Psi$.}, since there the role
of $\Psi$ is played by $e^{-\Phi}d e^\Phi$. Thus we have the puzzling result
that for superstrings, the tachyon is not necessary for describing physics
around the tachyon vacuum.

\subsection{$\psi_N$ piece}
\label{sec:psin}

To prove Sen's conjectures for the bosonic string it is necessary
to regulate the solution and subtract a mysterious term---the ``$\psi_N$ 
piece''---which vanishes in the Fock space\cite{Schnabl}. As we will see, 
a similar procedure is necessary for the superstring, but the story needs
some refinement.

The necessity of the $\psi_N$ piece can be understood from the requirement 
that the equations of motion hold in a sufficiently strong 
sense\cite{Okawa,Fuchs}. It is straightforward to prove the equations of motion
for \eq{super_sol} using the identities eqs.(\ref{eq:alg},\ref{eq:BRST}), 
but in the process we need to make the following assumption about the field 
$\frac{K}{1-F^2}$:
\begin{equation}\frac{KF^2}{1-F^2}=\frac{K}{1-F^2}-K
\label{eq:crux}\end{equation}
This apparently innocuous equation is where the subtleties with the 
$\psi_N$ piece come in.

The ``obvious'' solution to \eq{crux} is to define $\frac{K}{1-F^2}$ as a 
geometric series expansion,
\begin{equation}\frac{K}{1-F^2}=\lim_{N\to\infty}\sum_{n=0}^N K\Omega^n
\label{eq:naive}\end{equation}
Plugging in, one finds that \eq{crux} is satisfied up to a term,
\begin{equation}\lim_{N\to\infty}K\Omega^N\end{equation}
This actually vanishes in the Fock space as a power,
\begin{equation}\frac{1}{N^3}\end{equation}
but for the bosonic string this is not rapid enough to ensure the equations 
of motion hold when contracted with the solution\cite{Okawa,Fuchs}. For 
this purpose one needs \eq{crux} to hold up to $1/N^4\sim K^2\Omega^N$, which
requires the sliver state to be subtracted from the geometric sum. This is the
origin of the $\psi_N$ piece.

For more general purposes it may be useful to have a definition where \eq{crux}
is satisfied up to an arbitrary inverse power of $N$ in the Fock space. To 
see what the required corrections are, it is helpful to be 
more systematic. If we take \eq{crux} as given, one finds upon recursive 
substitution the identity:
\begin{equation}\frac{K}{1-F^2}=\sum_{n=0}^N K\Omega^n 
+ \left(\frac{K}{1-F^2}-K\right)\Omega^N\end{equation}
Now take the limit $N\to\infty$, and on the right hand side substitute the 
formal power series expansion,
\begin{equation}\frac{K}{1-F^2} = -\sum_{n=0}^\infty \frac{B_n}{n!}K^n
\end{equation}
in terms of Bernoulli numbers $B_n$. The result is the geometric expansion
\eq{naive} plus an infinite sequence of corrections involving powers of $K$ 
acting on $\Omega^N$. With a little Bernoulli arithmetic we can establish the 
following claim:

\bigskip
\noindent {\bf Claim:} The expression, 
\begin{equation}\frac{K}{1-F^2}=\lim_{N\to\infty}\left[\sum_{n=0}^N K\Omega^n
-\left(\sum_{k=0}^{A-2}\frac{B_k}{k!}K^k\,+K\right)\Omega^N\right]
\end{equation}
is a solution to \eq{crux} up to terms of order,
\begin{equation}\lim_{N\to\infty}K^A\Omega^N\sim\frac{1}{N^{2+A}}
\nonumber\end{equation}
with $A\geq 1$.
\bigskip

\noindent Though at the moment it is not obvious, as it happens we will need
$A=3$ for the superstring. Therefore we will take ($B_1=-\frac{1}{2}$),
\begin{equation}\frac{K}{1-F^2}=\lim_{N\to\infty}\left[\sum_{n=0}^N K\Omega^n
-\left(1+\frac{1}{2}K\right)\Omega^N\right]
\label{eq:reg0}\end{equation}
Plugging in to \eq{super_sol} gives a regulated expression of the superstring 
solution,
\begin{equation}\Psi = \lim_{N\to\infty}\left[\sum_{n=0}^N\psi_n'-\psi_N
-\frac{1}{2}\psi_N'\right]-\Gamma\label{eq:super_reg}\end{equation}
where,
\begin{eqnarray}\psi_n &=& Fc\Omega^n BcF\nonumber\\
\psi_n'&=&\frac{d}{dn}\psi_n = Fc\Omega^n KBcF\nonumber\\
\Gamma&=&FB\gamma^2F
\end{eqnarray}
The correction $-\frac{1}{2}\psi_N'$ is new and was not necessary for the 
bosonic string.

\section{Energy}
\label{sec:Energy}

Let us now calculate the energy. To prove 
Sen's conjecture we must demonstrate,
\begin{equation}E = -S(\Psi) = -\frac{1}{2\pi^2}\end{equation}
in the appropriate units\footnote{We normalize the basic correlator in the 
upper half plane,
$$\langle c\d c\d^2c(z_1)e^{-2\phi}(z_2)\rangle_{UHP}=-2$$
and set $\alpha'$, the open string coupling, and the spacetime volume factor 
to unity. Our normalization of the action agrees with Ohmori\cite{Ohmori2}.}.
Assuming the equations of motion and the regulated solution \eq{super_reg},
the energy breaks up into three terms:
\begin{equation}E = -\frac{1}{6}\Langle \Psi,Q_B\Psi\Rangle = E_B + E_\Gamma +
E_{\psi'}
\end{equation}
where $E_B$ is the contribution from the ``bosonic'' part of the solution,
$E_\Gamma$ is the contribution from the superstring correction term, and 
$E_{\psi'}$ is the contribution from the additional ``vanishing piece''
$-\frac{1}{2}\psi_N'$. Explicitly,
\begin{eqnarray}E_B &=& 
-\frac{1}{6}\lim_{N\to\infty}\left[\sum_{m,n=0}^N\Langle \psi_m',Q_B\psi_n'
\Rangle -2\sum_{m=0}^N\Langle \psi_m',Q_B\psi_N\Rangle+
\Langle \psi_N,Q_B\psi_N\Rangle\right]\nonumber\\
E_\Gamma &=& -\frac{1}{6}\lim_{N\to\infty}\left[\Langle \Gamma,Q_B\Gamma\Rangle
-2\sum_{m=0}^N\Langle\psi_m',Q_B\Gamma\Rangle
+2\Langle \psi_N,Q_B\Gamma\Rangle\right]
\nonumber\\
E_{\psi'}&=&-\frac{1}{6}\lim_{N\to\infty}
\left[\frac{1}{4}\Langle \psi_N',Q_B\psi_N'\Rangle 
-\sum_{m=0}^N\Langle \psi_m',Q_B\psi_N'\Rangle 
+ \Langle\psi_N,Q_B\psi_N'\Rangle +\Langle \psi_N',Q_B\Gamma\Rangle\right]
\nonumber\\ \end{eqnarray}

\begin{figure}
\begin{center}
\resizebox{2in}{1.7in}{\includegraphics{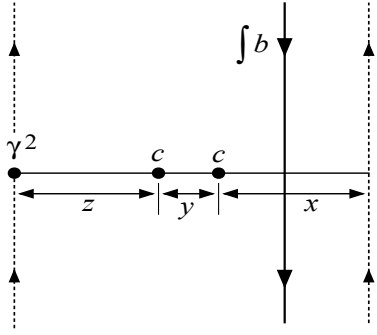}}
\end{center}
\caption{\label{fig:corr} Correlator \eq{corr1} on the cylinder. The dashed
vertical lines are identified, and the picture changing operator $Y_{-2}$ 
is inserted at the midpoint, $\pm i\infty$ in this coordinate system.}
\end{figure}

These expressions can be evaluated with knowledge of the inner products,
\begin{equation}\Langle \psi_m,Q_B\psi_n\Rangle\ \ \ \ 
\Langle \psi_n,Q_B\Gamma\Rangle\ \ \ \ \Langle \Gamma,Q_B\Gamma\Rangle
\end{equation}
An elementary computation with eqs.(\ref{eq:alg},\ref{eq:BRST}) reduces these
to a correlator on the cylinder (see figure \ref{fig:corr}):
\begin{equation}\Langle\Omega^x Bc\Omega^y c\Omega^z \gamma^2\Rangle
=\left\langle Y_{-2} \int_{i\infty}^{-i\infty}\frac{dw}{2\pi i}b(w) c(y+z)c(z)
\gamma^2(0)\right\rangle_{C_{x+y+z}}\label{eq:corr1}\end{equation}
We evaluate this in appendix \ref{app:A}, finding:
\begin{equation}\Langle\Omega^x Bc\Omega^y c\Omega^z \gamma^2\Rangle
=\frac{x+y+z}{2\pi^2}y\label{eq:corr2}\end{equation}
The inner products become,
\begin{eqnarray}\Langle \psi_m,Q_B\psi_n\Rangle &=& 
\frac{m+n+2}{\pi^2}\nonumber\\ 
\Langle \psi_m,Q_B \Gamma\Rangle &=& \frac{1}{\pi^2}\nonumber\\
\Langle \Gamma,Q_B\Gamma\Rangle &=& 0\label{eq:simple}
\end{eqnarray}
This result is vastly simpler than for the bosonic string, where 
$\langle \psi_m,Q_B\psi_n\rangle$ is a rather unwieldy expression
involving trigonometric functions\cite{Schnabl}. 

It is now extremely straightforward to evaluate the energy. Calculating the 
``bosonic'' contribution first,
\begin{eqnarray}E_B&=&-\frac{1}{6}\lim_{N\to\infty}\left[0-2\sum_{n=0}^N
\frac{1}{\pi^2}+\frac{2N+2}{\pi^2}\right]
=-\frac{1}{6}\lim_{N\to\infty}\left[-\frac{2(N+1)}{\pi^2}+\frac{2(N+1)}{\pi^2}
\right]\nonumber\\ &=&0\end{eqnarray}
We make a few comments. First, note that the ``Fock space'' contribution to 
the energy, from the double sum, vanishes identically because 
$\Langle\psi_m',Q_B\psi_n'\Rangle=0$. This is consistent with the 
expectation that the pure gauge solutions of Schnabl\cite{Schnabl} have 
vanishing energy. Second, note that,
$$\Langle \psi_N,Q_B\psi_N\Rangle$$
diverges linearly for large $N$, though fortunately this divergence cancels 
out of $E_B$. Thus, the inner product $\Langle \psi_N',Q_B\psi_N\Rangle$ 
will {\it a priori} make a finite contribution to the energy, which is why 
the subleading correction $-\frac{1}{2}\psi_N'$ to the $\psi_N$ piece is
important.

The contributions from $E_\Gamma$ and $E_\psi'$ are easily seen to be,
\begin{eqnarray}E_\Gamma &=& -\frac{1}{6}\lim_{N\to\infty}\left[0-0+2\cdot
\frac{1}{\pi^2}\right] = -\frac{1}{3\pi^2}\nonumber\\
E_{\psi'} &=& -\frac{1}{6}\lim_{N\to\infty}\left[0 - 0
+\frac{1}{\pi^2}+0\right]= -\frac{1}{6\pi^2}
\end{eqnarray}
Adding everything up,
\begin{equation}E = 0-\frac{1}{3\pi^2}-\frac{1}{6\pi^2} = -\frac{1}{2\pi^2}
\end{equation}
recovering the expected vacuum energy.

We can also prove that the equations of motion are satisfied when contracted
with the solution. Though this essentially follows from our previous 
discussion in section \ref{sec:psin}, it is worthwhile to check. We need to 
evaluate the cubic term,
\begin{equation}\Langle \Psi,\Psi*\Psi\Rangle\end{equation}
Since the picture changing insertion $Y_{-2}$ has $\phi$-momentum $-4$ and we 
need $\phi$-momentum $-2$ to get a nonvanishing correlator, the only 
nonvanishing contributions to the cubic term involve two $\psi_m$s and one 
$\Gamma$. Furthermore, because the correlator \eq{corr2} is linear in $x,z$, 
the contributions have at most one $\psi_m'$. Multiplying everything out, one 
finds the nonvanishing terms are,
\begin{equation}\Langle \Psi^3\Rangle = 3\lim_{N\to\infty}
\left[\sum_{m=0}^N\Langle \Gamma\psi_m'\psi_N\Rangle
+\sum_{m=0}^N\Langle \Gamma\psi_N\psi_m'\Rangle -\Langle \Gamma\psi_N^2\Rangle
-\frac{1}{2}\Langle \Gamma\psi_N'\psi_N\Rangle-\frac{1}{2}\Langle \Gamma
\psi_N\psi_N'\Rangle
\right]
\end{equation}
Calculating the inner product,
\begin{eqnarray}\Langle \Gamma\psi_m\psi_n\Rangle &=& 
\Langle \Omega^{m+1}Bc\Omega c\Omega^{n+1}\gamma^2\Rangle\nonumber\\
&=& \frac{m+n+3}{2\pi^2}
\end{eqnarray}
we find,
\begin{eqnarray}\Langle \Psi^3\Rangle &=& 3\lim_{N\to\infty}
\left[2\sum_{n=0}^N\frac{1}{2\pi^2}-\frac{2N+3}{2\pi^2}-\frac{1}{2\pi^2}
\right]\nonumber\\
&=& 3\lim_{N\to\infty}
\left[\frac{2N+2}{2\pi^2}-\frac{2N+3}{2\pi^2}-\frac{1}{2\pi^2}
\right]\nonumber\\
&=& -\frac{3}{\pi^2} 
\end{eqnarray}
and we have already calculated,
\begin{equation}\Langle \Psi,Q_B\Psi\Rangle= 6\cdot\frac{1}{2\pi^2}
=\frac{3}{\pi^2}
\end{equation}
proving the equations of motion are satisfied.

One interesting feature of these proofs is that they work even at finite $N$,
that is, the limit $N\to\infty$ was not necessary. Of course, at finite $N$ 
we do not really have a solution, but this can be fixed up by replacing our 
regulated expression \eq{reg0} with,
\begin{equation}\frac{K}{1-F^2}=\sum_{n=0}^N K\Omega^n
-\left(\sum_{k=0}^\infty\frac{B_k}{k!}K^k\,+K\right)\Omega^N
\end{equation}
For finite $N$ substituting the Bernoulli power series is somewhat formal, but 
interestingly for $N=0$ we recover the solution written in the $\L_0$ level 
expansion\footnote{In the current notation, a state $F\phi F$ has $\L_0$
eigenvalue $h$ if the operator insertion corresponding to the field $\phi$
has scaling dimension $h$ in the cylinder coordinate system. $K,B,c,\gamma^2$
have dimension $1,1,-1,-1$ respectively. Hence, for 
example, $FcK^nBcF$ has $\L_0$ eigenvalues $n-1$ and $FB\gamma^2F$ has 
$\L_0$ eigenvalue $0$.}:
\begin{eqnarray}\Psi &=& 
-\sum_{n=0}^\infty \frac{B_n}{n!}FcK^nBcF - FB\gamma^2F\nonumber\\
&=& -\sum_{n=0}^\infty \frac{B_n}{n!}
\left.\frac{d^n}{d\alpha^n}\right|_0\psi_\alpha -\Gamma\end{eqnarray}
Therefore, the fact that our calculation works independent of $N$ implies that
we have indirectly proven the energy in the $\L_0$ level expansion as well. 
If we like, we can repeat the proof in the new notation:
\begin{eqnarray}E &=& -\frac{1}{6}\Langle \Psi,Q_B\Psi\Rangle\nonumber\\
&=& -\frac{1}{6}\sum_{m,n=0}^\infty \frac{B_mB_n}{m!n!}
\left.\frac{\d^m}{\d\alpha^m}\right|_0\left.\frac{\d^n}{\d\beta^n}\right|_0
\Langle \psi_\alpha,Q_B\psi_\beta\Rangle
-\frac{1}{3}\sum_{m=0}^\infty \frac{B_m}{m!}
\left.\frac{\d^m}{\d\alpha^m}\right|_0\Langle \psi_\alpha,Q_B\Gamma\Rangle
\nonumber\\ &=& -\frac{1}{6}\left[\left(\frac{B_0}{0!}\right)^2
\Langle\psi_0,Q_B\psi_0\Rangle+2\frac{B_0B_1}{0!1!}
\Langle\psi_0',Q_B\psi_0\Rangle
\right]-\frac{1}{3}\frac{B_0}{0!}\Langle \psi_0,\Gamma\Rangle\end{eqnarray}
where in the last step we used the fact that second and higher derivatives 
of the inner products vanish. Plugging in \eq{simple} and $B_1=-\frac{1}{2}$,
\begin{equation}E=-\frac{1}{6}\left(\frac{2}{\pi^2}-\frac{1}{\pi^2}\right)
-\frac{1}{3\pi^2}=-\frac{1}{2\pi^2}
\end{equation}
For the bosonic string, evaluating the energy in the $\L_0$ level expansion
gives a very complicated asymptotic series, though the series can be resummed 
numerically to give a good approximation to the brane tension\cite{Schnabl}. 
One advantage of this derivation is that we do not need to regulate the 
solution or worry about subtracting the correct $\psi_N$ piece; these 
subtleties are implicitly taken care of in the $\L_0$ level expansion.

\section{Conclusion}

In this paper we have given a remarkably simple proof of Sen's conjectures 
in cubic superstring field theory. From an analytic perspective the solution
appears to be as regular as Schnabl's solution for the bosonic 
string. From the perspective of the level expansion the situation 
is unclear. Given the Siegel gauge results (see figure \ref{fig:level}) we 
expect convergence to be irregular\footnote{In fact, Ohmori\cite{Ohmori2} 
searched for, but failed to find a GSO$(+)$ vacuum out to level (2,6).
We find this worrisome, but it may be hard to identify a vacuum in level 
truncation because the cubic coupling $\Langle c,c,c\Rangle$ vanishes by 
$\phi$-momentum conservation. Therefore the auxiliary tachyon must depend
on interactions with higher level fields to generate a minimum.}, but perhaps 
the situation will improve at sufficiently high level. 

Given the intrinsic uncertainties of the cubic theory, it is highly desirable
to construct an analytic vacuum solution to Berkovits's WZW-type 
superstring field theory. Following the philosophy of 
refs.\cite{Ohmori1,Erler3}, it is not difficult to construct formal solutions
once an appropriate solution for the cubic equations of motion is 
known---in fact, one such vacuum solution has already been proposed in 
ref.\cite{Fuchs_super}. However, the GSO$(+)$ Berkovits solutions we have found
seem to be singular in the $\L_0$ level expansion. We suspect that an analytic
vacuum solution in the Berkovits theory will have to involve the 
GSO$(-)$ sector in some nontrivial way. This may be expected from level 
truncation analysis, which shows a smooth double-well potential with minima 
for the tachyon at finite expectation value.

The biggest puzzle presented by our solution is the absence of any component
in the GSO$(-)$ sector. This brings up three apparent paradoxes: 

\begin{figure}
\begin{center}
\resizebox{2in}{1.7in}{\includegraphics{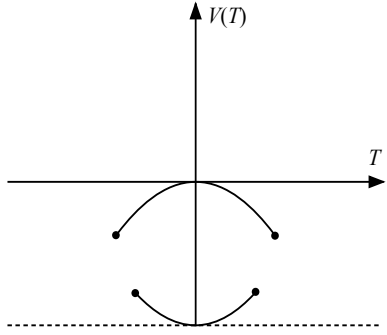}}
\end{center}
\caption{\label{fig:super_pot} Conjectured form of the tachyon potential in
Schnabl gauge for the cubic superstring. The minimum on the lower branch 
represents our analytic solution \eq{super_sol}. The dots at the edge of 
the curves represent points on the tachyon potential where Schnabl gauge 
breaks down.}
\end{figure}

\begin{enumerate}
\item There is an expectation, which so far has been unchallenged, that open 
string field theory on a particular brane system only describes classical 
solutions which are accessible via tachyon condensation. A BPS brane carries 
a conserved topological charge, so there is no means for it to decay to the 
vacuum, by tachyon condensation or otherwise. Thus it appears that the cubic
superstring has ``too many'' solutions. 

\item The intuitive picture of tachyon condensation suggests that the 
tachyon should roll off the top of the potential and come to rest at the 
vacuum with finite expectation value. However, the story here must be 
different. It seems possible that the tachyon potential in Schnabl gauge hits 
a singularity before a stable minimum is reached, and the GSO$(+)$ solution 
lies on a disconnected branch directly below the unstable maximum 
(see figure \ref{fig:super_pot}). This scenario is made
a more plausable from the level $\frac{5}{2}$ results of 
Raeymaekers\cite{Raeymaekers}. It is also circumstantially supported by the 
(somewhat mysterious) late time rolling tachyon limit of 
Ellwood\cite{Ellwood}. For the bosonic string, there seems to be a sense in 
which the late time behavior of the rolling tachyon solution\cite{marginal}
approaches Schnabl's solution. However, a similar limit for the 
superstring\cite{Erler3,Yuji} fails to yield a well-defined expression, 
suggesting a vacuum solution in Schnabl gauge with nonvanishing GSO$(-)$ 
sector may not exist.

\item The third paradox comes from supersymmetry. Since the perturbative 
vacuum on the BPS brane has unbroken supersymmetry, one would not 
expect to find a state in the theory with lower energy.
\end{enumerate}
It would be very interesting to gain concrete insight into these 
puzzles.

The author would like to thank I.Ellwood, E.Fuchs, J. Raeymaekers, M. Schnabl 
and A. Sen for useful conversations. The author also thanks D.Gross and the 
KITP in Santa Barbara for hospitality while some of this work was in progress. 
This work is supported by the National Science Foundation under Grant No.NSF 
PHY05-51164 and by the Department of Atomic Energy, Government of India.

\begin{appendix}

\section{Correlator}
\label{app:A}

In this appendix we derive the correlator \eq{corr2} used to derive the inner 
products and energy in section \ref{sec:Energy}. We start with:
\begin{eqnarray}\Langle\Omega^x Bc\Omega^y c\Omega^z \gamma^2\Rangle
&=&\left\langle Y_{-2} \int_{i\infty}^{-i\infty}\frac{dw}{2\pi i}b(w) 
c(z_1)c(z_2)\gamma^2(0)\right\rangle_{C_L}\nonumber\\
L &=& x+y+z\ \ \ \ z_1=y+z\ \ \ \ z_2=z\end{eqnarray}
as shown in figure \ref{fig:corr}. To simplify the $b$ ghost insertion,
we use the trick of Okawa\cite{Okawa}. We introduce a linear function on 
the cylinder $(z)_\delta$ with a branch cut at $\mathrm{Re}(z)=\delta$, and 
write the $b$ insertion as a contour integral around this branch cut:
\begin{equation}\int_{i\infty}^{-i\infty} \frac{dw}{2\pi i} b(z)= 
\frac{1}{L}\oint_{\mathrm{Re}(z)=\delta} \frac{dw}{2\pi i} (w)_\delta b(w)
\end{equation}
The factor of $1/L$ is necessary because the discontinuity has height $L$ 
for a cylinder of circumference $L$. We then deform the contour away from the 
branch cut to encircle the $c$ insertions inside the correlator:
\begin{eqnarray}\Langle\Omega^x Bc\Omega^y c\Omega^z \gamma^2\Rangle
&=&-\frac{1}{L}
\left\langle Y_{-2} \left(\oint_{c\mathrm{s}}\frac{dw}{2\pi i}wb(w) 
c(z_1)c(z_2)\right)\gamma^2(0)\right\rangle_{C_L}\nonumber\\
&=& -\frac{z_1}{L}\langle Y_{-2} c(z_2)\gamma^2(0)\rangle_{C_L}
+\frac{z_2}{L}\langle Y_{-2} c(z_1)
\gamma^2(0)\rangle_{C_L}
\end{eqnarray}
The remaining correlator can be evaluated by mapping back to the upper 
half plane and performing the necessary contractions. The answer is,
\begin{equation}\langle Y_{-2} c(z)\gamma^2(0)\rangle_{C_L}=
-\frac{L^2}{2\pi^2}\label{eq:simp_cor}\end{equation}
The simplicity of this result is responsible for all of the drastic 
simplifications of the energy calculation. We can see that the basic structure
is correct by inspection: The factor of $L^2$ is necessary because the 
insertions have total conformal dimension $-2$. Furthermore, the result
must be independent of $z$ because $c(z)$ and 
$\gamma^2(0)$ only have contractions with the picture changing 
operators at $\pm i\infty$, and by cylindrical symmetry these contractions 
are independent of the absolute or relative positions of these operators.
Thus,
\begin{eqnarray}\Langle\Omega^x Bc\Omega^y c\Omega^z \gamma^2\Rangle
&=& \frac{L}{2\pi^2}(z_1-z_2)\nonumber\\
&=& \frac{x+y+z}{2\pi^2} y\end{eqnarray}
reproducing \eq{corr2}.

\end{appendix}

\end{document}